\documentclass[twocolumn,showpacs,preprintnumbers,amsmath,amssymb]{revtex4}
\usepackage{tabularx,graphicx}

\usepackage{color}
\usepackage{hyperref}
\hypersetup{
    colorlinks=true,
    linkcolor=blue,
    filecolor=blue,      
    urlcolor=blue,
}

\usepackage{color}

\usepackage{ulem}   

\begin{document}

\newcommand{\beq}{\begin{equation}}
\newcommand{\eeq}{\end{equation}}
\newcommand{\beqn}{\begin{eqnarray}}
\newcommand{\eeqn}{\end{eqnarray}}
\newcommand{\bmath}{\begin{subequations}}
\newcommand{\emath}{\end{subequations}}
\newcommand{\bra}[1]{\langle #1|}
\newcommand{\ket}[1]{|#1\rangle}

\title{Absence of evidence of superconductivity in  sulfur hydride in optical reflectance experiments}

\author{J. E. Hirsch$^{a}$  and F. Marsiglio$^{b}$ }
\address{$^{a}$Department of Physics, University of California, San Diego,
La Jolla, CA 92093-0319\\
$^{b}$Department of Physics, University of Alberta, Edmonton,
Alberta, Canada T6G 2E1}

\maketitle

 Capitani and coworkers \cite{t} reported that infrared optical reflectance measurements provided  evidence for a superconducting transition in sulfur hydride  \cite{sh3} under  150 GPa pressure, and  that the transition is driven by the
 electron-phonon interaction. 
Here we argue that the measured data did not provide evidence that the system undergoes a transition
to a superconducting state, nor do the data support any role of phonons in driving a transition. Rather, the data are consistent with  the system remaining in the normal state 
down to  temperature  50K, the lowest temperature measured in the experiment.
This calls into further question \cite{hm2345,huang} the generally accepted view \cite{citations} that sulfur hydride under pressure is a high
temperature superconductor.

We requested all the raw data used in Ref.~\cite{t} from the corresponding authors, and obtained them for temperatures 50 K, 100 K, 150 K, 240 K and 320 K. The paper also shows data for temperatures 200 K, 130 K and 170 K; we reiterated our request for those raw data but did not receive them.

Figure~1 shows raw data for H$_3$S (main frame). This data represents an intensity detected at the spectrometer, and is comprised
of a combination of blackbody source radiation and reflectance off the sample and diamond anvil apparatus \cite{ttpc}.
Also shown in the bottom right inset for comparison is the (presumably raw) reflectance data from a
cuprate sample \cite{ybco}. In an attempt
to remove non-intrinsic  signals from the H$_3$S data, the authors of Ref.~\cite{t} plotted 
ratios of intensity at different temperatures. This is shown in the inset to the immediate right of the main figure, while the
corresponding procedure for the cuprates is given at the far right side (top).

 It is clear that at low temperature, the cuprate reflectance flattens and approaches unity on the scale given by the superconducting
energy gap $2 \Delta_0$, estimated as $2 \Delta_0 \approx 350 \ {\rm cm}^{-1}=43.4$ meV for that case. All cuprate reflectance curves,
below and above $T_c$, monotonically increase with decreasing frequency/energy (except for some sharp dips
due to phonons) and approach unity at low
frequency/energy, and for a given frequency, the reflectance monotonically decreases as the temperature increases.
In the inset showing the ratio of the reflectance for the two lowest temperatures for the cuprate (far right at the top), a peak is present, consistent
with the expectation given in Fig.~1b of \cite{t}. One of these curves is at a temperature above $T_c$, and also shows a gap,
consistent with the original authors' interpretation of their normal state data in terms of a pseudogap.
The contrast with the complicated pattern in the inset on the left showing  reflectance ratios for H$_3$S  is striking.

         \begin{figure} [t]
    \resizebox{8.5cm}{!}{\includegraphics[width=6cm]{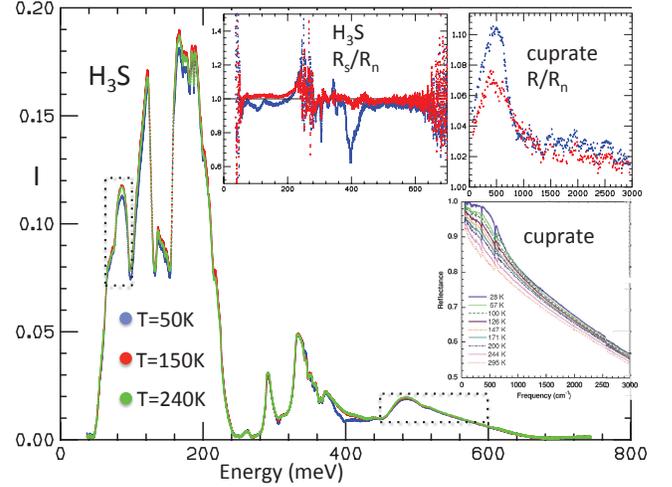}} 
 \caption { Raw data for measured reflectance of sulfur hydride underlying the results presented in Ref.~\cite{t}. 
 The dotted rectangles mark the regions of energy shown in Figs.~2b and 3b of Ref.~\cite{t}.
 The lower right  inset shows reflectance data for a cuprate superconductor, with $T_c=59$ K from Ref.~\cite{ybco}. To convert 
 frequency (cm$^{-1}$) to meV, multiply by $0.124$. The upper left and right insets show reflectance ratios for 
 H$_3$S and for the cuprate respectively: $R_n$ is for $T=240K$ ($T=200K$) for the left (right) upper inset; $R_s$ in the upper left inset is for 
 temperature $T=50K$ (blue) and $T=150K$ (red). $R$ for the upper right inset is for $T=28K$ (blue)  and $T=67K$ (red). }
 \label{figure1}
 \end{figure} 
 
The usual technique of evaporating gold on a free-
standing sample to obtain an absolute reflectance is
not feasible here \cite{ttpc}. It is clear that
the main intensity in the body of Fig.~1 has large contributions from sources
{\it other than} the H$_3$S sample. Using Eq.~(3) in the Supplementary Material of Ref. \cite{t} we can relate the ratio of the sample
reflectance in the superconducting and normal state to the ratio of intensities at the same two temperatures:
\beq
\label{eq1}
{R_s \over R_n} = {I_s \over I_n} - {a \over R_n}\left( 1 - {I_s \over I_n} \right),
\eeq
where the parameter $a=0.13$ if we use the values for $R_c$ and $\alpha$ provided in Ref.~\cite{t}. Since the second term
on the right-hand-side of Eq.~(\ref{eq1}) is at most of order 2\% and tends to correct the ratio towards unity, we approximate the
ratio of reflectances by the ratio of intensities.
It is clear from comparison of the upper left and right insets that 
the  reflectance ratios 
in the case of H$_3$S do not indicate
superconductivity.

To zoom in on this issue, in Fig.~2 we show the intensity raw data and ratios in the energy range 65 meV-100 meV that Ref.~\cite{t} focuses
on to infer that H$_3$S undergoes a superconducting transition. The solid curves on the right panel are 
taken from Fig.~2b of \cite{t}; however, to match the measured data we had to shift the blue curve given in Ref.~\cite{t} $downward$ by 0.042 and
the red curve $upward$ by 0.016. Note that these are significant shifts on the scale shown. 
The fact that the low temperature reflectance ratio is smaller than the high temperature
reflectance ratio, opposite to what Ref.~\cite{t} shows, directly contradicts the interpretation that these features are {\it ``in good agreement with the theoretical gap
structure''} as is claimed in Ref.~\cite{t}.  The left panel of Fig.~2 shows that both the temperature and energy dependence of the
measured curves are dominated by effects not coming from the sample of interest, H$_3$S.

          \begin{figure} [t]
  \resizebox{8.5cm}{!}{\includegraphics[width=6cm]{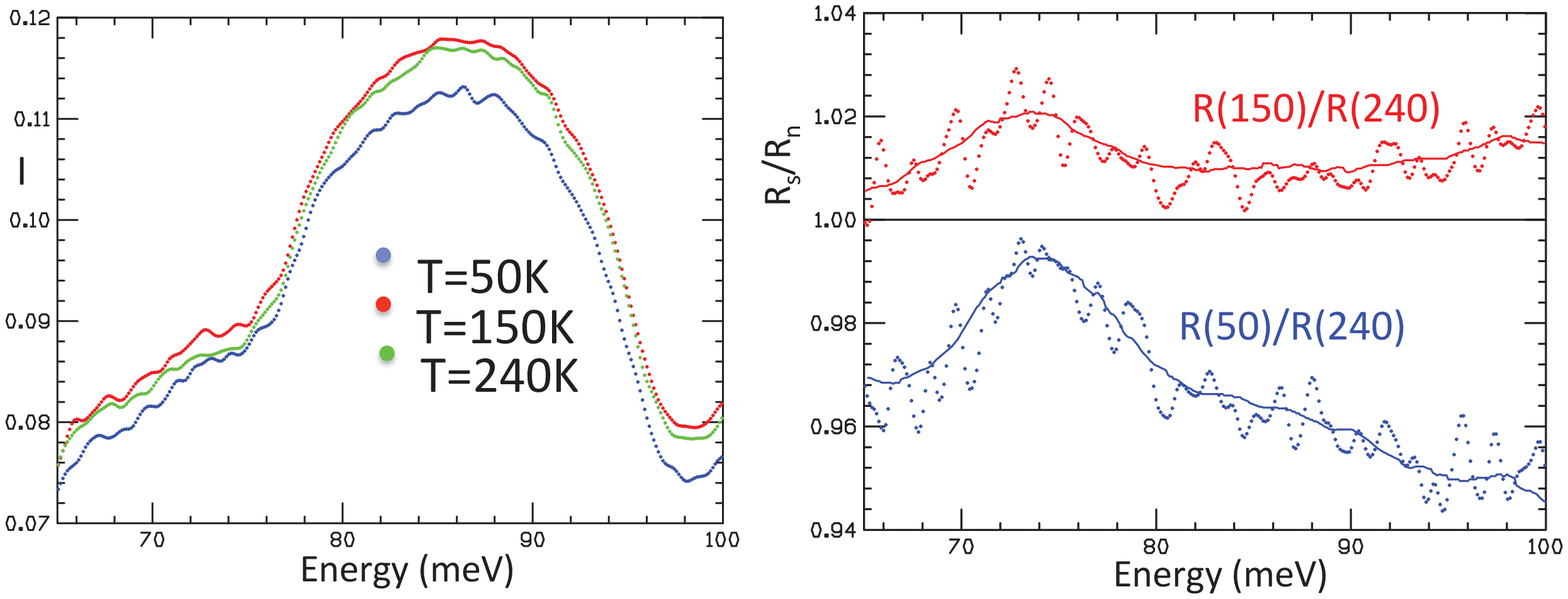}} 
 \caption {Left panel: intensity raw data. Right panel: reflectance ratios. The lines show the curves displayed in Fig.~2b of 
 ref. \cite{t}, shifted down (up) by 0.042 (0.016) for the blue (red) curve.   }
 \label{figure1}
 \end{figure}

Ref. \cite{t} also shows reflectance ratios in their Fig.~3b in the energy range 450 meV to 600 meV, reproduced here on the
right panel of Fig.~3,  and they claim that their positive
slope and temperature dependence is in agreement with theoretical calculations   and ``{\it demonstrates that
H$_3$S is an Eliashberg superconductor driven by the electron-phonon interaction with strong coupling to
high phonons of order of $200$ meV}''. For comparison, on the left panel of Fig.~3 we plot the indicated reflectance ratios obtained from the raw data that we obtained. 
The red points on the left panel should be the same as the light blue curve on the right panel, but they are not. The blue
points on the left panel, for a lower temperature, should have a steeper positive slope as the dark blue
curve on the right panel, but they do not. Instead, both sets of points on the left panel show  a rather noisy and flat 
behavior qualitatively different from the behavior in the right panel in the same energy range,
450 meV to 600 meV. For energies  below 450 meV, not shown on the right panel, the low temperature (blue) 
curve on the left panel shows a sudden sharp drop which is $not$ predicted by the theory used in 
Ref.~\cite{t} (see   the dashed blue line in Fig. 3a of \cite{t}). It has been suggested that it is due to ice depositing on the
diamond surface at low temperature \cite{ttpc}.
We argue that the measured data shown in the left panel of Fig.~3, which markedly differ from the curves shown
in the right panel,  provide no support to the hypothesis that
Ref.~\cite{t} claims the right panel of Fig.~3 {\it ``demonstrates''}.

            \begin{figure} [t]
 \resizebox{8.5cm}{!}{\includegraphics[width=6cm]{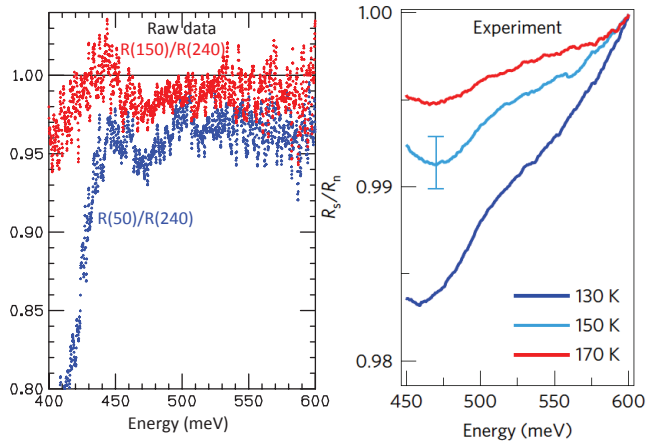}} 
 \caption {Left panel: reflectance ratios from raw data. Right panel: curves from Fig.~3b of Ref.~\cite{t}.  }
 \label{figure1}
 \end{figure} 
 
Returning to Fig.~2b in Ref.~\cite{t}, the figure caption reads 
  ``{\it The blue and red solid curves indicate the measured $R_s(T)/R_n$
for T=50 K and 150 K, respectively with $R_n$ measured at 200 K.}''   Given that the energy dependence of the curves
in Ref.~\cite{t}  follows 
closely the behavior
we obtain from the raw data, as shown in our Fig.~2 right panel, we conclude that the paper has a typo and
$R_n$ was measured at 240 K. However, it is clear from our Fig.~2 that the curves shown in the paper were not
`{\it `the measured $R_s(T)/R_n$''}, but rather  the measured $R_s(T)/R_n$ \textit{\textbf{shifted}} as indicated in the figure
caption of our Fig. 2. This shift is crucial since it led to a  temperature dependence claimed in the paper 
exactly \textit{\textbf{opposite}} to what the measured data showed. The authors concluded from their Fig.~2b  that  {\it ``The intensity of this feature
and its temperature dependence are in good agreement with the theoretical
behaviour of the superconducting gap''}.
Instead, we conclude that the data plotted in Fig.~2b
of Ref.~\cite{t}, rather than  being {\it ``the measured $R_s(T)/R_n$''}  that the figure caption stated,  resulted from alteration
of the measured data in unexplained and  unjustified ways, and that  the measured data provide {\it no support} for
the existence of a superconducting gap.

  In summary, we argue that Ref.~\cite{t} presents a misleading picture. Data for carefully selected small energy windows
  were chosen,  ratios of measured quantities rather than absolute values were plotted, and measured data were altered in
  unexplained ways, to arrive at the published data, in order to hide the complicated and
  unexpected behavior of the measured data shown in Fig.~1. The purpose of these selections and
  alterations was clearly  to provide apparent support to the a priori $assumption$ that H$_3$S is a {\it high temperature conventional superconductor.}  
  As a result, the reader  of Ref.~\cite{t} is left with the impression that
  these optical measurements provided (i) confirmation that sulfur hydride is a superconductor and (ii) confirmation that the
  electron-phonon interaction drives its superconductivity. 
  Instead, we have shown here that in fact the measured data  provided support for neither. The reality is that the measured data have major contributions from
unknown sources that render them incapable of supplying $any$ information about the physics of H$_3$S.

\begin{acknowledgments}
  The authors are grateful to Tom Timusk and Pascale Roy for providing the raw data and
  additional information and  for helpful discussions.
  FM was supported in part by the Natural Sciences and Engineering
Research Council of Canada (NSERC) and by an MIF from the Province of Alberta.

\end{acknowledgments}

   \end{document}